\documentclass[aps,twocolumn,showpacs,superscriptaddress]{revtex4}
\usepackage{amsmath}
\usepackage{mathrsfs}
\usepackage{amssymb}
\usepackage{amsfonts}
\usepackage{graphicx}

\begin{document}

\title{Spin Squeezing, Negative Correlations, and Concurrence in the Quantum
Kicked Top Model}
\author{Xiaoqian Wang}
\affiliation{Department of Physics, Changchun University of Science and Technology,
Changchun 130022, P. R. China}
\affiliation{Zhejiang Institute of Modern Physics, Department of Physics, Zhejiang
University, Hangzhou 310027 P. R. China}
\author{Jian Ma}
\affiliation{Zhejiang Institute of Modern Physics, Department of Physics, Zhejiang
University, Hangzhou 310027 P. R. China}
\author{Lijun Song}
\affiliation{School of Science, Changchun University, Changchun 130022, P. R. China}
\author{Xihe Zhang}
\affiliation{Department of Physics, Changchun University of Science and Technology,
Changchun 130022, P. R. China}
\author{Xiaoguang Wang}
\email{xgwang@zimp.zju.edu.cn}
\affiliation{Zhejiang Institute of Modern Physics, Department of Physics, Zhejiang
University, Hangzhou 310027 P. R. China}
\email{xgwang@zimp.zju.edu.cn}

\begin{abstract}
We study spin squeezing, negative correlations, and concurrence in the
quantum kicked top model. We prove that the spin squeezing and negative
correlations are equivalent for spin systems with only symmetric Dicke
states populated. We numerically analyze spin squeezing parameters and
concurrence in this model, and find that the maximal spin squeezing
direction, which refers to the minimal pairwise correlation direction, is
strongly influenced by quantum chaos. Entanglement (spin squeezing) sudden
death and sudden birth occur alternatively for the periodic and
quasi-periodic cases, while only entanglement (spin squeezing) sudden death
is found for the chaotic case.
\end{abstract}

\maketitle

\section{Introduction}

Quantum entanglement is one of the central concepts in quantum information
theory, and can be viewed as a physical resource for quantum
information\thinspace \cite{Einstein,Amico,Horodecki}. Characterizing
entanglement is still an open question for many-body system, however, for
two-qubit case, the problem is solved by using concurrence which is defined
by Wootters\thinspace \cite{Wootters0,Wootters2001}. In Refs.\thinspace \cite%
{Amico,Horodecki,Wineland,Kitagawa,Toth,Alekseev,Soensen,Pu,Messikh,wang2009,Yin,Orgikh}%
, it was found that spin squeezing has a close relation with entanglement,
and spin squeezing parameters can be used as entanglement witnesses. The
definitions of spin squeezing parameters are not unique, and the most
popular parameters are $\xi _{KU}^{2}$, given by Kitagawa and Ueda\thinspace
\cite{Kitagawa}, and $\xi _{W}^{2}$, given by Wineland \textit{et
al\thinspace }\cite{Wineland}. The parameter $\xi _{KU}^{2}$ was shown to be
related to pairwise correlations and concurrence\textit{\thinspace }\cite%
{Kitagawa,Orgikh,Messikh,wang2009,Yin}, while the parameter $\xi _{W}^{2}$
is related to many-body entanglement\textit{\thinspace }\cite%
{Wineland,Messikh}. There is another spin squeezing parameter proposed by S%
\o rensen \textit{et al\thinspace }\cite{Soensen} as a measure of many-body
entanglement which can be considered as a generalization of parameter $\xi
_{W}^{2}$\textit{\thinspace }\cite{Messikh}. Recently, a new spin squeezing
parameter $\xi _{T}^{2}$ was proposed, and it was found that if $\xi
_{T}^{2}<1$, the system is entangled\textit{\thinspace }\cite{Toth,wang2009}.

Reference~\cite{Orgikh} found that spin squeezing parameter $\xi _{KU}^{2}$
has close relation with pairwise correlation. Here, we consider the
following pairwise correlation
\begin{equation}
\mathcal{C}_{i\vec{n},j\vec{n}}=\left\langle \sigma _{i\vec{n}}\sigma _{j%
\vec{n}}\right\rangle -\left\langle \sigma _{i\vec{n}}\right\rangle
\left\langle \sigma _{j\vec{n}}\right\rangle ,
\end{equation}%
in the $\vec{n}$ direction, where $\sigma _{i\vec{n}}=\vec{\sigma}_{i}\cdot
\vec{n}$, with $\vec{\sigma}_{i}$ being the Pauli matrix of the $i$-th qubit
and $\vec{n}$ being the normalized direction vector. The above definition
can be rewritten as
\begin{equation}
\mathcal{C}_{i\vec{n},j\vec{n}}=\vec{n}\mathbb{C}\vec{n}^{T},
\end{equation}%
where $\mathbb{C}$ is the pairwise correlation matrix whose elements are
given by
\begin{equation}
\mathbb{C}_{i\alpha ,j\beta }=\left\langle \sigma _{i\alpha }\sigma _{j\beta
}\right\rangle -\left\langle \sigma _{i\alpha }\right\rangle \left\langle
\sigma _{j\beta }\right\rangle ,\text{ \ }\alpha ,\beta =x,y,z.
\end{equation}

In Ref.~\cite{Orgikh}, it was found that, if the system has exchange
symmetry, the pairwise correlation and the spin squeezing parameter $\xi
_{KU}^{2}$ have the following relation,%
\begin{equation}
\mathcal{C}_{\vec{n}_{\perp },\vec{n}_{\perp }}=\frac{\xi _{KU}^{2}-1}{N-1},
\label{pc_ku}
\end{equation}%
where indices $i$, $j$ were omitted due to exchange symmetry, $\vec{n}%
_{\perp }$ is the direction which is perpendicular to the mean spin
direction $\vec{n}$, and $N$ is the number of qubits. In this paper, we find
that the parameter $\xi _{T}^{2}$ is related to the minimal pairwise
correlation for systems with only symmetric Dicke states populated\thinspace~%
\cite{Dicke,SDicke}. The relation is given by
\begin{equation}
\mathcal{C}_{\min }=\min_{\vec{n}}\mathcal{C}_{\vec{n},\vec{n}}=\frac{\xi
_{T}^{2}-1}{N-1},  \label{pc_tu}
\end{equation}%
where the minimization is over arbitrary direction $\vec{n}$. Thus, for
systems with only symmetric Dicke states populated, negative correlations $%
\left( \mathcal{C}_{\min }<0\right) $ are equivalent to the spin squeezing $%
\left( \xi _{T}^{2}<1\right) $.

Many works have been devoted to understand spin squeezing, quantum
entanglement, and quantum chaos\textit{\thinspace }\cite%
{Gorin,Jacquod,Wang,Wang0,Wang2003,Weinstein0,Ghose,Stamatiou,Song,Furuya,Petijean,Soensen,Song0}%
. Here, quantum chaos\thinspace \cite{Gutzwiller,Hans,Peres} mainly focuses
on the researches of quantum characteristics of a quantum system whose
classical correspondence exhibits chaos. In Refs.\textit{\thinspace }\cite%
{Song,Song0}, the authors used the parameter $\xi _{KU}^{2}$ as an efficient
signature of quantum chaos. They also studied the relations between $\xi
_{KU}^{2}$ and concurrence, and found that $\xi _{KU}^{2}$ does not match
concurrence well. In this paper, we use the new spin squeezing parameter $%
\xi _{T}^{2}$ to characterize quantum chaos, and study the dynamical
evolutions of concurrence and spin squeezing parameters $\xi _{KU}^{2}$ and $%
\xi _{T}^{2}$. The quantum kicked top (QKT) model\textit{\thinspace }\cite%
{Arecchi,Haake,ARIANO,Ghose,Stamatiou,Song} is a typical model that exhibits
quantum chaos, and its chaotic behaviors were demonstrated experimentally by
using linear entropy\textit{\thinspace }\cite{Chaudhury}. In this paper we
find that the concurrence decreases abruptly and non-smoothly to zero in a
finite time in the QKT model, and this phenomenon is called entanglement
sudden death (ESD)\thinspace \cite{Miranowicz}, which has been widely
studied both theoretically\thinspace \cite%
{Yu2004,Yu2006,Cavalcanti,Huang,Almeida,Sainz,Qasimi,Yamanoto,Salles,Yu2009,Faria,Pablo,Weinstein2009}
and experimentally\textit{\thinspace }\cite{Salles,Almeida} in decoherence
dynamics. Similar to the definition of ESD, there is entanglement sudden
birth (ESB)\thinspace \cite{Ficek,Lopez,Mazzola,Amri,Zhang}, which is a
sudden feature in the temporal creation of entanglement in a dissipative
evolution of interacting qubits. In Ref.\thinspace \cite{wang2009}, the
authors found that spin squeezing sudden death (SSSD) may occur due to
decoherence. In this paper, we want to study these sudden features of
entanglement and spin squeezing in the QKT model.

This paper is organized as follows. In Sec.\textit{\thinspace }II, we first
introduce the definitions of spin squeezing parameters, negative
correlations, and concurrence, then give the relationship between the spin
squeezing parameter $\xi _{T}^{2}$ and the minimal pairwise correlation $%
\mathcal{C}_{\min }$. In Sec.\textit{\thinspace }III, we introduce the QKT
model and its classical correspondence. In Sec.\textit{\thinspace }IV, we
analyze the quantum chaos of the QKT model by means of the dynamics of spin
squeezing and entanglement. In Sec.\textit{\thinspace }V, we study the
influences of chaos on ESD, ESB, SSSD and spin squeezing sudden birth
(SSSB). The conclusions are given in Sec.\thinspace VI.

\section{Spin squeezing, negative correlations, and concurrence}

\subsection{Definitions of Spin squeezing parameters}

The spin squeezing parameters~are useful tools to detect the quantum
entanglement~\cite{Wineland,Kitagawa,Toth,Alekseev,Soensen,Pu}. Here, we
consider an ensemble of $N$ spin-$1/2$ particles described by the collective
angular momentum operators
\begin{equation}
J_{\alpha }=\frac{1}{2}\sum_{k=1}^{N}\sigma _{k\alpha },\text{ \ \ }\alpha
=x,y,z.  \label{collective}
\end{equation}%
The Dicke states can be expressed as $J_{+}^{n}|1\rangle ^{\otimes N}$,
where $|1\rangle $ is the spin down state and $J_{\pm }=J_{x}\pm iJ_{y}$. We
mainly study the following two types of spin squeezing parameters $\xi
_{KU}^{2}$ and $\xi _{T}^{2}$. The parameter $\xi _{KU}^{2}$ is defined as~%
\cite{Kitagawa}
\begin{equation}
\xi _{KU}^{2}=\frac{4\left( \triangle J_{\vec{n}_{\perp }}\right) _{\min
}^{2}}{N},  \label{sp_ku}
\end{equation}%
where $\vec{n}_{\perp }$ denotes the direction which is perpendicular to the
mean spin direction $\vec{n}=\big\langle\vec{J}\big\rangle/\big\vert\vec{J}%
\big\vert$, and $\left( \triangle J_{\vec{n}_{\perp }}\right) _{\min }^{2}$
is the minimal value of the variance $\left( \Delta J\right) ^{2}$ in the $%
\vec{n}_{\perp }$-direction. The spin squeezing parameter $\xi _{KU}^{2}$
can be written in an explicit form as\thinspace \cite{Wang2003}%
\begin{eqnarray}
\xi _{KU}^{2} &=&\frac{2}{N}\bigg{[}\langle J_{\vec{n}_{1}}^{2}+J_{\vec{n}%
_{2}}^{2}\rangle   \notag \\
&&-\sqrt{(\langle J_{\vec{n}_{1}}^{2}-J_{\vec{n}_{2}}^{2}\rangle
)^{2}+\langle \lbrack J_{\vec{n}_{1}},J_{\vec{n}_{2}}]_{+}\rangle ^{2}}\bigg{]},
\label{sp_ku1}
\end{eqnarray}%
where $\vec{n}_{1}$ and $\vec{n}_{2}$ are two orthogonal directions which
are perpendicular to the mean spin direction, and $\left[ A,B\right]
_{+}=AB+BA$.

The spin squeezing parameter $\xi _{T}^{2}$ is defined as~\cite%
{Toth,wang2009}%
\begin{equation}
\xi _{T}^{2}=\frac{\lambda _{\min }}{\big\langle \vec{J}^{2}\big\rangle -%
\frac{N}{2}},  \label{sp_toth}
\end{equation}%
where $\lambda _{\min }$ is the minimal eigenvalue of the real symmetric
matrix
\begin{equation}
\Gamma =\left( N-1\right) \gamma +\mathbf{C},  \label{pc-sp}
\end{equation}%
where $\mathbf{C}$ is the correlation matrix of which the matrix elements
are
\begin{equation}
\mathbf{C}_{\alpha \beta }=\frac{1}{2}\left\langle J_{\alpha }J_{\beta
}+J_{\beta }J_{\alpha }\right\rangle ,\text{ \ \ }\alpha ,\beta =x,y,z,
\end{equation}%
The covariance matrix $\gamma $ is given as%
\begin{equation}
\gamma _{\alpha \beta }=\mathbf{C}_{\alpha \beta }-\left\langle J_{\alpha
}\right\rangle \left\langle J_{\beta }\right\rangle .
\end{equation}

Below, for the sake of simplicity, we assume the mean spin direction $\vec{n}
$ is along the $z$-axis. And then we can write the matrix $\Gamma $ in an
explicit form as%
\begin{equation}
\Gamma =\left(
\begin{array}{cc}
\Gamma _{\vec{n}_{\perp }} & \vec{B}^{T} \\
\vec{B} & \Gamma _{\vec{n}}%
\end{array}%
\right) ,  \label{Gamma}
\end{equation}%
where $\Gamma _{\vec{n}_{\perp }}$ is a $2\times 2$ matrix%
\begin{equation}
\Gamma _{\vec{n}_{\perp }}=\left(
\begin{array}{cc}
N\big\langle J_{x}^{2}\big\rangle & \frac{N}{2}\big\langle \left[ J_{x},J_{y}%
\right] _{+}\big\rangle \\
\frac{N}{2}\big\langle \left[ J_{x},J_{y}\right] _{+}\big\rangle & N%
\big\langle J_{y}^{2}\big\rangle%
\end{array}%
\right) ,
\end{equation}%
and
\begin{eqnarray}
\Gamma _{\vec{n}} &=&N\left( \bigtriangleup J_{z}\right) ^{2}+\big\langle %
J_{z}\big\rangle ^{2}, \\
\vec{B} &=&\left( \frac{N}{2}\big\langle \left[ J_{z},J_{x}\right] _{+}%
\big\rangle ,\frac{N}{2}\big\langle \left[ J_{z},J_{y}\right] _{+}%
\big\rangle \right) .
\end{eqnarray}%
Note that $\vec{B}$ is a $1\times 2$ vector. According to Eq.\thinspace (\ref%
{sp_ku1}), the parameter $\xi _{KU}^{2}$ is equal to the minimal eigenvalue
of the matrix%
\begin{equation}
\tilde{\Gamma}=\frac{4}{N^{2}}\Gamma _{\vec{n}_{\perp }}=\frac{4}{N^{2}}%
\left(
\begin{array}{cc}
N\big\langle J_{x}^{2}\big\rangle & \frac{N}{2}\big\langle \left[ J_{x},J_{y}%
\right] _{+}\big\rangle \\
\frac{N}{2}\big\langle \left[ J_{x},J_{y}\right] _{+}\big\rangle & N%
\big\langle J_{y}^{2}\big\rangle%
\end{array}%
\right) .
\end{equation}%
In analogy to the relation between $\xi _{KU}^{2}$ and $\Gamma _{\vec{n}%
_{\perp }}$, we define a spin squeezing parameter%
\begin{equation}
\xi _{\vec{n}}^{2}=\frac{4}{N^{2}}\Gamma _{\vec{n}}=\frac{4}{N^{2}}\left[
N\left( \bigtriangleup J_{z}\right) ^{2}+\left\langle J_{z}\right\rangle ^{2}%
\right] ,  \label{zetann}
\end{equation}%
which characterizes the spin squeezing along the mean spin direction. Here
we consider the case of $\big\langle \vec{J}^{2}\big\rangle =\frac{N}{2}%
\left( \frac{N}{2}+1\right) $. According to the Rayleigh-Ritz
theorem\thinspace \cite{Roger}, the minimal eigenvalue of $\Gamma $ is less
or equal to that of $\Gamma _{\vec{n}_{\perp }}$, thus we have $\xi
_{T}^{2}\leq \xi _{KU}^{2}$.

\subsection{Relations between spin squeezing and the minimal pairwise
correlation}

Now, we discuss the relation between the spin squeezing parameter $\xi
_{T}^{2}$ and the minimal pairwise correlation $\mathcal{C}_{\min }$. Here,
we consider the system with only symmetric Dicke states populated, which has
exchange symmetry and $\big\langle \vec{J}^{2}\big\rangle =\frac{N}{2}\left(
\frac{N}{2}+1\right) $. In this case, we have the following relations,
\begin{align}
\big\langle J_{\alpha }^{2}\big\rangle & =\frac{N}{4}+\frac{N\left(
N-1\right) }{4}\big\langle \sigma _{1\alpha }\sigma _{2\alpha }\big\rangle ,
\\
\big\langle J_{-}^{2}\big\rangle & =N\left( N-1\right) \big\langle \sigma
_{1-}\sigma _{2-}\big\rangle , \\
\big\langle \left[ J_{\alpha },J_{\beta }\right] _{+}\big\rangle & =\frac{%
N\left( N-1\right) }{4}\big\langle \left[ \sigma _{1\alpha },\sigma _{2\beta
}\right] _{+}\big\rangle ,\text{ }\left( \alpha \neq \beta \right).
\end{align}%
Thus, with the above relations, the matrix $\Gamma $, as shown in
Eq.\thinspace (\ref{pc-sp}), can be rewritten as%
\begin{align}
\Gamma _{\alpha ,\beta }& =\frac{N}{2}\big\langle \left[ J_{\alpha
},J_{\beta }\right] _{+}\big\rangle -\left( N-1\right) \big\langle J_{\alpha
}\big\rangle \big\langle J_{\beta }\big\rangle  \notag \\
& =\frac{N^{2}\left( N-1\right) }{4}\left( \big\langle \sigma _{\alpha
}\sigma _{\beta }\big\rangle -\big\langle \sigma _{\alpha }\big\rangle %
\big\langle \sigma _{\beta }\big\rangle \right)  \notag \\
& =\frac{N^{2}\left( N-1\right) }{4}\mathbb{C}_{\alpha ,\beta },\text{ \ \ }%
\left( \alpha \neq \beta \right) ;
\end{align}%
\begin{align}
\Gamma _{\alpha ,\alpha }& =N\big\langle J_{a}^{2}\big\rangle -\left(
N-1\right) \big\langle J_{\alpha }\big\rangle ^{2}  \notag \\
& =\frac{N^{2}\left( N-1\right) }{4}\left( \big\langle \sigma _{1\alpha
}\sigma _{2\alpha }\big\rangle -\big\langle \sigma _{\alpha }\big\rangle %
^{2}\right) -\frac{N^{2}}{4}  \notag \\
& =\frac{N^{2}\left( N-1\right) }{4}\mathbb{C}_{\alpha ,\alpha }-\frac{N^{2}%
}{4}.
\end{align}%
The relation between the matrix $\Gamma $ and the pairwise correlation
matrix $\mathbb{C}$ can be written as%
\begin{equation}
\mathbb{C}=\frac{4\Gamma }{N^{2}\left( N-1\right) }-\frac{\mathbb{I}}{\left(
N-1\right) }.
\end{equation}%
where $\mathbb{I}$ is a $3\times 3$ identity matrix. Thus, the matrix $%
\mathbb{C}$ and $\Gamma $ can be diagonalized with the same unitary
transformation. So
\begin{equation}
\mathcal{C}_{\min }=\frac{4\lambda _{\min }}{N^{2}\left( N-1\right) }-\frac{1%
}{N-1},
\end{equation}%
where $\mathcal{C}_{\min }$ and $\lambda _{\min }$ are the minimal
eigenvalues of the matrix $\mathbb{C}$ and $\Gamma $, respectively. Here the
spin squeezing parameter $\xi _{T}^{2}$ can be written as $\xi
_{T}^{2}=4\lambda _{\min }/N^{2}$ since $\big\langle \vec{J}^{2}\big\rangle =%
\frac{N}{2}\left( \frac{N}{2}+1\right) $. The relation between the minimal
pairwise correlation and the parameter $\xi _{T}^{2}$ can be written as
\begin{equation}
\mathcal{C}_{\min }=\min_{\vec{n}}\mathcal{C}_{\vec{n},\vec{n}}=\frac{\xi
_{T}^{2}-1}{\left( N-1\right) }.  \label{neg_spqt}
\end{equation}%
Therefore, negative correlation ($\mathcal{C}_{\min }<0$) is equivalent to
spin squeezing ($\xi _{T}^{2}<1$). From Refs.\textit{\thinspace }\cite%
{Toth,wang2009}, $\xi _{T}^{2}<1$ is also a criterion of entanglement, so it
reveals that there are close relations among negative correlation, spin
squeezing, and entanglement. If the minimal pairwise correlation is in the
plane which is perpendicular to the mean spin direction, the above relation
will degenerate to Eq.\thinspace (\ref{pc_ku}).

\subsection{Relations between spin squeezing and concurrence}

Here, we briefly introduce the relations between spin squeezing and
concurrence. The entanglement between a pair of spin-$1/2$ particles is
quantified by the concurrence $C$\thinspace \cite{Wootters0,Wootters2001},
which is defined as%
\begin{equation}
C=\max \left\{ 0,\lambda _{1}-\lambda _{2}-\lambda _{3}-\lambda _{4}\right\}
,  \label{conc}
\end{equation}%
where the quantities $\lambda _{i}$ are the square roots of the eigenvalues
of the matrix $\rho _{12}\left( \sigma _{1y}\otimes \sigma _{2y}\right) \rho
_{12}^{\ast }\left( \sigma _{1y}\otimes \sigma _{2y}\right) $ in descending
order, $\rho _{12}^{\ast }$ is the complex conjugate of $\rho _{12}$. For
spin states with parity, it was found that\thinspace \cite{Wang2003}, when $%
\xi _{KU}^{2}<1$, the relation between $\xi _{KU}^{2}$ and $C$ is
\begin{equation}
\xi _{KU}^{2}=1-\left( N-1\right) C.  \label{sq_con}
\end{equation}%
In Refs.\thinspace \cite{Yin}, the authors found that for systems with
parity and exchange symmetry, when concurrence $C>0$, the spin squeezing
parameter $\xi _{T}^{2}<1$, and vice versa.

\section{Quantum Kicked Top}

Now we introduce the QKT model. Consider an ensemble of $N$ spin-$1/2$
particles, the QKT Hamiltonian reads\thinspace \cite{Arecchi,Haake,ARIANO}
\begin{equation}
H=\frac{\kappa }{2j\tau }J_{z}^{2}+pJ_{y}\sum_{n=-\infty }^{\infty }\delta
\left( t-n\tau \right) ,  \label{qkt_hamiltonian}
\end{equation}%
where $j=N/2$, and $J_{\alpha }$ $\left( \alpha =x,y,z\right) $ are angular
momentum operators that obey the commutation relations $\left[ J_{\alpha
},J_{\beta }\right] =i\hbar \varepsilon _{\alpha \beta \gamma }J_{\gamma }$,
where $\varepsilon _{\alpha \beta \gamma }$ is the Levi-Civita symbol. The
first term of Eq.\thinspace (\ref{qkt_hamiltonian}) describes a nonlinear
precession around the $z$-axis with strength $\kappa $, and the second term
describes the kicks\ around the $y$-axis with strength $p$, separated by a
period $\tau $. In the below, we set $p=\pi /2$ and $\tau =1$, and the
magnitude $\big\langle \vec{J}^{2}\big\rangle =j\left( j+1\right) $ is a
constant of the motion.

\begin{figure}[tbp]
\begin{center}
\includegraphics[
clip, width=9cm ]{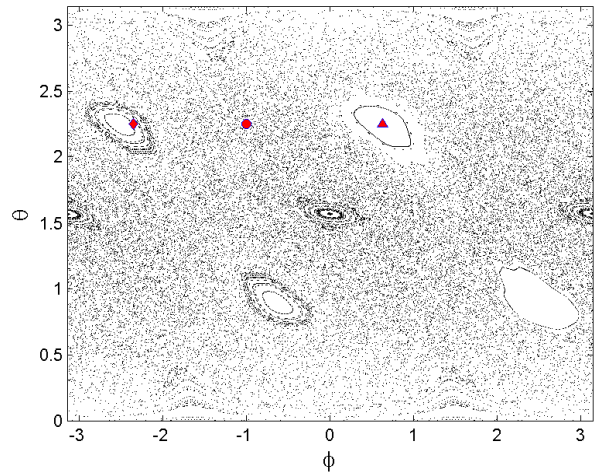}
\end{center}
\caption{(Color online) Stroboscopic phase-space map of the classical kicked
top with $\protect\kappa =3$ and $p=\protect\pi /2.$ we choose 200 random
initial states and each evolves 200 kicks. A triangle at $(\protect\theta ,%
\protect\phi )=(2.25,0.63)$ is at the fixed point, namely in the periodic
region. A diamond at $\left( \protect\theta ,\protect\phi \right) =\left(
2.25,-2.35\right) $ is in the quasi-periodic region. A circle at $\left(
\protect\theta ,\protect\phi \right) =\left( 2.25,-1\right) $ is in the
chaotic region.}
\label{classical_map}
\end{figure}

Now, we study the classical corresponding of the QKT. The evolutions of the
expectation values of the angular momentum operators are%
\begin{equation}
\left\langle J_{a}\right\rangle _{n+1}=\left\langle U^{\dagger }J_{\alpha
}U\right\rangle _{n},  \label{evolution}
\end{equation}%
where $U$ is the Floquet operator describing the unitary evolution for each
kick,%
\begin{equation}
U=\exp \left( -\frac{i\kappa }{2j}J_{z}^{2}\right) \exp \left(
-ipJ_{y}\right) .  \label{Floquet}
\end{equation}%
To study the quantum chaos of the QKT, we should first analyze its classical
corresponding, which is obtained in the classical limit, i.e. $j\rightarrow
\infty $.\ For convenience, we use the following three quantities
\begin{equation}
X=\frac{\left\langle J_{x}\right\rangle }{j},\text{ \ }Y=\frac{\left\langle
J_{y}\right\rangle }{j},\text{ \ }Z=\frac{\left\langle J_{z}\right\rangle }{j%
},  \label{JXYZ}
\end{equation}%
when $j\rightarrow \infty $, these three variables become%
\begin{equation}
\left( X,Y,Z\right) =\left( \sin \theta \cos \phi ,\sin \theta \sin \phi
,\cos \phi \right) ,  \label{XYZ}
\end{equation}%
where $\theta $ is the polar angle and $\phi $ is the azimuthal angle.
Therefore, $\left( X,Y,Z\right) $ represents a point on the Bloch sphere
with radius $r=1$. In the classical limit, we can factorize the products of
the mean values of the angular momentum operators as
\begin{equation}
\frac{\left\langle J_{x}J_{y}\right\rangle }{j^{2}}=XY\text{.}  \label{JXY}
\end{equation}%
By substituting Eqs.\thinspace (\ref{Floquet}), (\ref{JXYZ}), and (\ref{JXY}%
) into Eq.\thinspace (\ref{evolution}), we can derive the classical
equations of motions as\thinspace \cite{Haake}%
\begin{equation}
\left[
\begin{array}{c}
X \\
Y \\
Z%
\end{array}%
\right] _{n+1}=\left[
\begin{array}{c}
Z\cos \left( \kappa X\right) +Y\sin \left( \kappa X\right) \\
-Z\sin \left( \kappa X\right) +Y\cos \left( \kappa X\right) \\
-X%
\end{array}%
\right] _{n}.
\end{equation}%
Therefore, in the classical limit, the dynamics of the QKT is governed by
the above equation, and by using Eq.\thinspace (\ref{XYZ}), we plot the
stroboscopic dynamics of the classical variables $\theta $ and $\phi $ in
Fig.\thinspace \ref{classical_map}. Each point represents a state of $%
(X,Y,Z) $ in the phase space. In this plot, we choose $\kappa =3$, and thus
there are periodic, quasi-periodic, and chaotic regions in the phase space.

The quantum evolution of the QKT is studied when $j$ is finite, and the
chaos is indicated by the stroboscopic plot shown in Fig.\thinspace \ref%
{classical_map}. Thus, to make connection between quantum and classical
evolutions, the initial state should be chosen as coherent spin state
(CSS)\thinspace \cite{Arecchi,Haake,ARIANO}, which can be viewed as a most
classical state%
\begin{equation}  \label{eq36}
|\theta ,\phi \rangle =R\left( \theta ,\phi \right) |j,j\rangle=R\left(
\theta ,\phi \right) |1\rangle^{\otimes N} ,
\end{equation}%
where $|j,j\rangle $ is the eigenstate of $J_{z}$ with the eigenvalue $j$,
and the rotation operator is defined as
\begin{equation}  \label{eq37}
R\left( \theta ,\phi \right) =\exp \left\{ i\theta \left[ J_{x}\sin \phi
-J_{y}\cos \phi \right] \right\} ,
\end{equation}%
where $0\leq \theta \leq \pi $ and $0\leq \phi \leq 2\pi $. The expectation
values of the angular momentums on this state are given by
\begin{equation}
\left\langle \mathbf{J}\right\rangle =\left( \left\langle J_{x}\right\rangle
,\left\langle J_{y}\right\rangle ,\left\langle J_{z}\right\rangle \right)
=j\left( \sin \theta \cos \phi ,\sin \theta \sin \phi ,\cos \theta \right) ,
\end{equation}%
which is the same as Eq.\thinspace (\ref{XYZ}). This is the reason that we
choose CSS to be the initial state. Since CSS can be regarded as a classical
state, and it is a product state, thus there is no correlations between
qubits. Substituting Eq.~(\ref{collective}) into Eq.~(\ref{eq37}), we can
see that the operator $R$ can be written as a direct product of $N$
operators acting independently on each qubit. And note that $|j,j\rangle $
is a product state, and thus the CSS is also a product state. In the
classical limit, the initial classical state is indeed a CSS, and during the
evolution, the state is still classical. However, in the quantum case, the
factorization\thinspace (\ref{JXY}) is not valid, since quantum correlations
are created during the evolution, and thus the state is no longer a CSS.
Such quantum correlations are characterized below by spin squeezing and
concurrence.

\section{Dynamics of spin squeezing and concurrence in the QKT model}

\begin{figure}[tbp]
\includegraphics[width=8cm,clip]{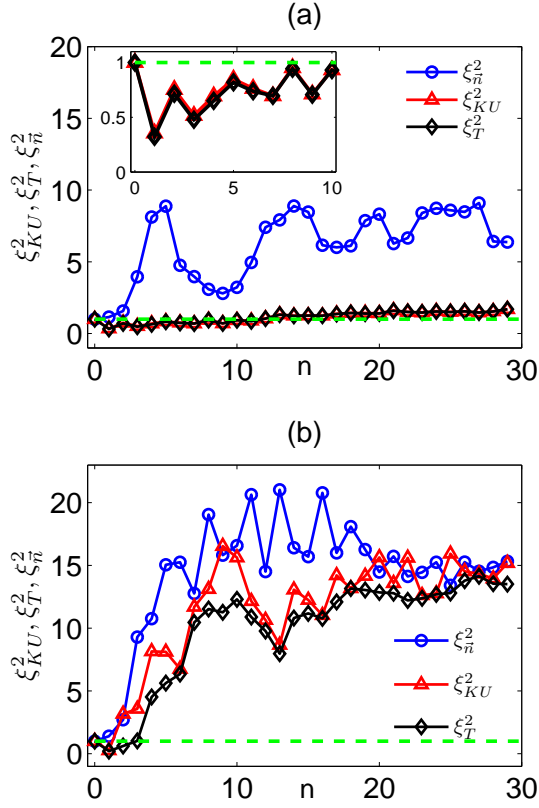}
\caption{(Color online) Dynamical evolutions of the parameters $\protect\xi %
_{T}^{2},\protect\xi _{KU}^{2}$, and $\protect\xi _{\vec{n}}^{2}$ with
different initial states. We choose $N=50.$ In the periodic region (a), $%
\left( \protect\theta ,\protect\phi \right) =\left( 2.25,0.63\right) ,$ $%
\protect\xi _{KU}^{2}$ and $\protect\xi _{T}^{2}$ are almost the same, while
$\protect\xi _{\vec{n}}^{2}$ is very large. This implies that, the maximal
spin squeezing is around the $\vec{n}_{\perp }$-direction. However, in the
chaotic region (b), $\left( \protect\theta ,\protect\phi \right) =\left(
2.25,-1\right) ,$ the differences between $\protect\xi _{KU}^{2}$ and $%
\protect\xi _{T}^{2}$ are obvious, that means, the maximal spin squeezing is
not restricted to the $\vec{n}_{\perp }$-direction. The dashed line
corresponds to $1$. }
\label{spinsqu}
\end{figure}

At first, we use spin squeezing to characterize quantum correlations and
quantum chaos. According to the discussions of Sec.\thinspace II, the spin
squeezing parameter $\xi _{T}^{2}$ characterizes the minimal pairwise
correlation, the parameter $\xi _{KU}^{2}$ characterizes the minimal
pairwise correlation in the plane which is perpendicular to the mean spin
direction $\vec{n}$, and the parameter $\xi _{\vec{n}}^{2}$ is the pairwise
correlation along $\vec{n}$-direction. The numerical results of quantum
evolutions of spin squeezing parameters $\xi _{T}^{2}$, $\xi _{KU}^{2}$, and
$\xi _{\vec{n}}^{2} $ are given in Fig.\thinspace \ref{spinsqu}, and we find
that the maximal spin squeezing directions are strongly influenced by
quantum chaos.

Here, we analyze these three spin squeezing parameters in the periodic
region, as shown in Fig.\thinspace \ref{spinsqu}(a). It can be seen that the
spin squeezing parameter $\xi _{KU}^{2}$ is much smaller than $\xi _{\vec{n}%
}^{2}$, but there are very slight differences between $\xi _{KU}^{2}$ and $%
\xi _{T}^{2}$, thus the maximal spin squeezing, which refer to the minimal
pairwise correlation, is around the $\vec{n}_{\perp }$-direction.

In the chaotic case, as shown in Fig.\thinspace \ref{spinsqu} (b), we can
find that the parameter $\xi _{T}^{2}$ is much smaller than both $\xi
_{KU}^{2}$ and $\xi _{\vec{n}}^{2}$ at some time, that means the maximal
spin squeezing is along neither the $\vec{n}_{\perp }$-direction nor the $%
\vec{n}$-direction. Therefore, the directions of maximal spin squeezing are
around the $\vec{n}_{\perp }$-direction in the periodic case, while they are
not restricted to the $\vec{n}_{\perp }$-direction in the chaotic case.

\begin{figure}[tbp]
\includegraphics[clip,width=9cm]{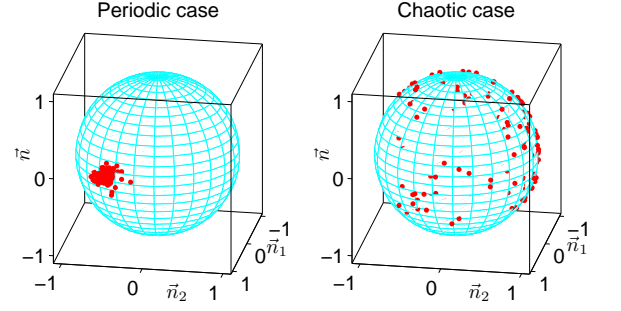}
\caption{(Color online) Directions of the maximal spin squeezing. The dots
on the uint sphere represent the end points of the directions of maximal
spin squeezing. And the axis label $\vec{n}$ denotes the mean spin
direction, $\vec{n}_{1}$ and $\vec{n}_{2}$ are two orthogonal directions
which are perpendicular to $\vec{n}$. In the left plot, the initial state is
in the periodic region, with $\left( \protect\theta ,\protect\phi \right)
=\left( 2.25,0.63\right) $. And in the right plot, the initial state is in
the chaotic region, with $\left( \protect\theta ,\protect\phi \right)
=\left( 2.25,-1\right) $. }
\label{spq_dir}
\end{figure}

The directions of the maximal spin squeezing are calculated in
Fig.\thinspace \ref{spq_dir}. The $\vec{n}$-axis in Fig.\thinspace \ref%
{spq_dir} is the mean spin direction\thinspace \cite{Song}. It can be
written in the spherical coordinate as
\begin{equation}
\vec{n}=\left( \sin \theta _{0}\cos \phi _{0},\sin \theta _{0}\sin \phi
_{0},\cos \theta _{0}\right) ,
\end{equation}%
where $\theta _{0}$ and $\phi _{0}$ are the polar and azimuthal angles,
respectively. They are calculated as
\begin{eqnarray}
\theta _{0} &=&\arccos \left( \big\langle J_{z}\big\rangle /\big\vert \vec{J}%
\big\vert \right) ,  \notag \\
\phi _{0} &=&\left\{
\begin{array}{c}
\arccos \left( \frac{\left\langle J_{x}\right\rangle }{\left\vert \vec{J}%
\right \vert \sin \theta _{0}}\right) \text{ \ \ \ \ \ \ \ if }\left\langle
J_{y}\right\rangle >0, \\
2\pi -\arccos \left( \frac{\left\langle J_{x}\right\rangle }{\left\vert \vec{%
J}\right\vert \sin \theta _{0}}\right) \text{ \ \ if }\left\langle
J_{y}\right\rangle \leq 0.%
\end{array}%
\right.
\end{eqnarray}%
where $\big\vert \vec{J}\big\vert =\sqrt{\big\langle J_{x}\big\rangle
^{2}+\big\langle J_{y}\big\rangle ^{2}+\big\langle J_{z}\big\rangle
^{2}}$. The other two axes in Fig.\thinspace \ref{spq_dir} are chosen as%
\begin{eqnarray}
\vec{n}_{1} &=&\left( -\cos \theta _{0}\cos \phi _{0},-\cos \theta _{0}\sin
\phi _{0},\sin \theta _{0}\right) ,  \notag \\
\vec{n}_{2} &=&\left( -\sin \phi _{0},\cos \phi _{0},0\right) .
\end{eqnarray}%
From Fig.\thinspace \ref{spq_dir}, it can be easily found that the
directions of the maximal spin squeezing are around the $\vec{n}_{\perp }$%
-direction in the periodic case. But in the chaotic case, the directions are
not concentrated in a certain direction. Thus the directions of maximal spin
squeezing are strongly influenced by quantum chaos.

\begin{figure}[tbp]
\includegraphics[clip,width=9cm]{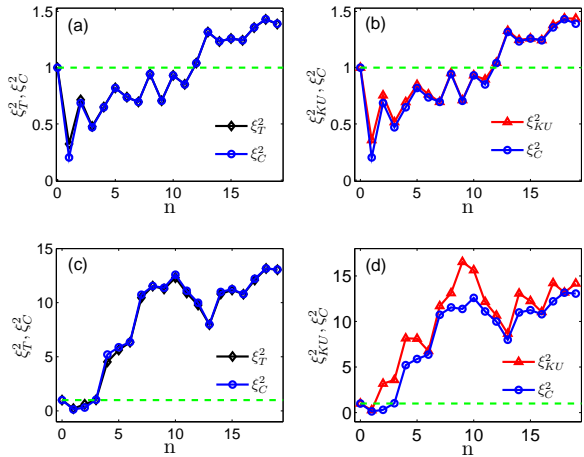}
\caption{(Color online) Evolutions of $\protect\xi _{T}^{2}$, $\protect\xi %
_{KU}^{2}$ and $\protect\xi _{C}^{2}$ with different initial states, and we
choose $N=50.$ In (a) and (b), the initial state is in the periodic region,
at $\left( \protect\theta ,\protect\phi \right) =\left( 2.25,0.63\right) $.
The values of $\protect\xi _{T}^{2}$ and $\protect\xi _{C}^{2}$ are almost
the same except for the first two kicks, but there are small differences
between $\protect\xi _{KU}^{2}$ and $\protect\xi _{C}^{2}$. In (c) and (d),
we choose $\left( \protect\theta ,\protect\phi \right) =\left(
2.25,-1\right) $, which is centered in the chaotic region. There are small
differences between $\protect\xi _{T}^{2}$ and $\protect\xi _{C}^{2}$, but
the differences between $\protect\xi _{KU}^{2}$ and $\protect\xi _{C}^{2}$
are large. The dashed line corresponds to $1$. }
\label{concurrence_spinsqu}
\end{figure}

Then, we use concurrence to character the quantum chaos, and give
comparisons among the spin squeezing parameters $\xi _{KU}^{2}$, $\xi
_{T}^{2} $, and concurrence. From Ref.\thinspace \cite{Wang2003}, when $\xi
_{KU}^{2}<1$, the relation between $\xi _{KU}^{2}$ and $C$ is shown as
Eq.\thinspace (\ref{sq_con}) for states with a fixed parity, and when $\xi
_{KU}^{2}\geq 1$, the relation does not hold. In the QKT model, there is not
a simple relation between spin squeezing and concurrence as the states here
are more general than those with a parity. In order to make a direct
comparison between spin squeezing parameter and concurrence, here, we
introduce a new quantity to describe the concurrence, the form is%
\begin{equation}
\xi _{C}^{2}=1-\left( N-1\right) C_{1},  \label{spin_con}
\end{equation}%
where
\begin{equation*}
C_{1}=\lambda _{1}-\lambda _{2}-\lambda _{3}-\lambda _{4},
\end{equation*}%
which can be got from the Eq.\thinspace (\ref{conc}) without the max
function, and when $\xi _{C}^{2}<1$, $C_{1}>0$, namely the state is pairwise
entangled. The numerical results of the dynamical evolutions of $\xi _{T}^{2}
$, $\xi _{KU}^{2}$, and $\xi _{C}^{2}$ in the QKT model are given in
Fig.\thinspace \ref{concurrence_spinsqu}.

In the periodic case, as shown in Fig.\thinspace \ref{concurrence_spinsqu}%
(a) and (b), we can see that, at the first two kicks, there are small
differences between $\xi _{T}^{2}$ $\left( \xi _{KU}^{2}\right) $ and $\xi
_{C}^{2}$. After the third kick, the parameters $\xi _{T}^{2}$ and $\xi
_{C}^{2}$ are nearly coincide, while there are also small differences
between $\xi _{KU}^{2}$ and $\xi _{C}^{2}$. We also note that when $\xi
_{C}^{2}<1$, both $\xi _{T}^{2}$ and $\xi _{KU}^{2}$ are smaller than $1$,
so both the two spin squeezing parameters can well describe the pairwise
entanglement.

In the chaotic case, as shown in Fig.\thinspace \ref{concurrence_spinsqu}(c)
and (d), we can find that there are very small differences between $\xi
_{T}^{2}$ and $\xi _{C}^{2}$, but the differences between $\xi _{KU}^{2}$
and $\xi _{C}^{2}$ are large. We also observe that when $\xi _{C}^{2}<1$, $%
\xi _{T}^{2}<1$, and vice versa. It means the spin squeezing parameter $\xi
_{T}^{2}$ may indicate the pairwise entanglement. At the second and third
kicks, the parameter $\xi _{C}^{2}<1$, $\xi _{KU}^{2}>1$, namely, the spin
squeezing parameter $\xi _{KU}^{2}$ is not a good quantity to describe the
pairwise entanglement in this chaotic case.

\section{Sudden death of entanglement and spin squeezing}

\begin{figure}[tbp]
\includegraphics[clip, width=9cm]{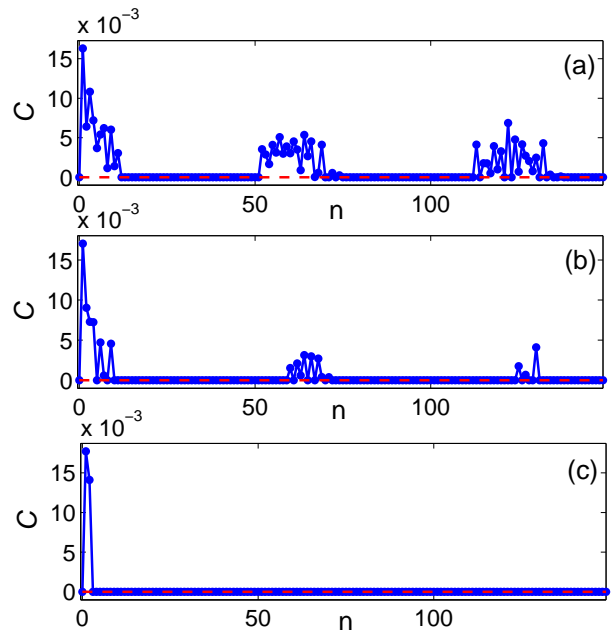}
\caption{(Color online) Dynamical evolutions of concurrence in the QKT model
for (a) the periodic case $\left( \protect\theta ,\protect\phi \right)
=\left( 2.25,0.63\right) $, (b) the quasi-periodic case $\left( \protect%
\theta ,\protect\phi \right) =\left( 2.25,-2.35\right)$, and (c) the chaotic
case $\left( \protect\theta ,\protect\phi \right) =\left( 2.25,-1\right) $.
Here, we choose $N=50$.}
\label{ESD-ESB}
\end{figure}

Recently, entanglement sudden death and sudden birth have received a lot of
attentions. In this paper, we find that both ESD and ESB occur in the QKT
model. Here we study the influences of quantum chaos on ESD and ESB. We
choose the initial states are in the periodic, quasi-periodic, and chaotic
regions, and the dynamics of concurrence are shown in Fig.\thinspace \ref%
{ESD-ESB}. Since the initial state is a CSS, there is no entanglement at
first, after the first kick, the pairwise entanglement ($C>0$) is produced.
As time evolves, the concurrence decreases to zero, and remains for a period
of time. As shown in Fig.\thinspace \ref{ESD-ESB}~(c), we can see that, when
the initial state is in the chaotic region, there is no entanglement again.
However, we observe ESB when the initial states are in the periodic and
quasi-periodic regions, as shown in Fig.\thinspace \ref{ESD-ESB}~(a) and
(b). The whole length of the time intervals for zero entanglement in the
periodic case is shorter than that in the quasi-periodic case.

\begin{figure}[tbp]
\includegraphics[clip, width=9cm]{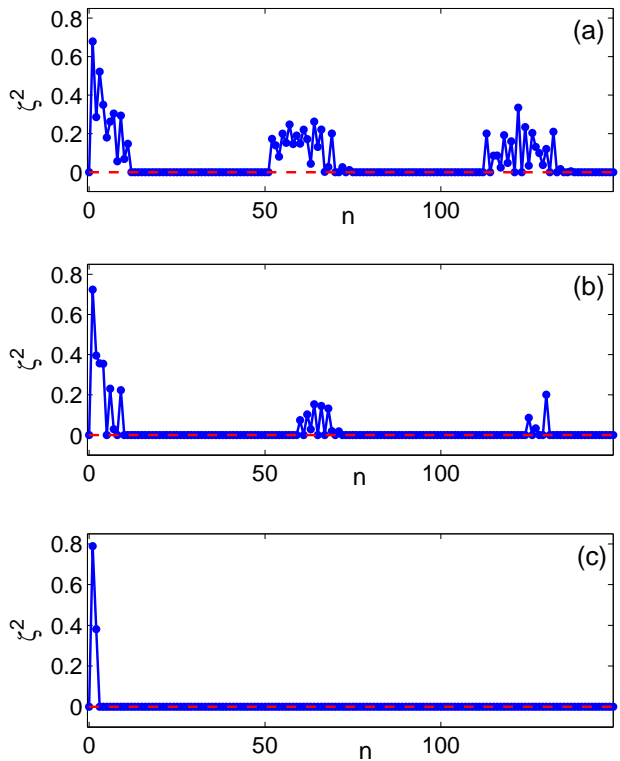}
\caption{(Color online) Dynamical evolutions of spin squeezing in the QKT
model with (a) the periodic case $\left( \protect\theta ,\protect\phi %
\right) =\left( 2.25,0.63\right) $, (b) the quasi-periodic case $\left(
\protect\theta ,\protect\phi \right) =\left( 2.25,-2.35\right) ,$ and (c)
the chaotic case $\left( \protect\theta ,\protect\phi \right) =\left(
2.25,-1\right) $. Here, we choose $N=50$.}
\label{SSSD-SSSB}
\end{figure}

Similar to concurrence, we also find SSSD and SSSB in the QKT model. In
Ref.\thinspace \cite{wang2009}, the authors introduced a quantity to
describe the spin squeezing, the form is
\begin{equation}
\zeta ^{2}=\max \left\{ 0,1-\xi _{T}^{2}\right\} ,
\end{equation}
therefore, if there is no spin squeezing, i.e. $\xi _{T}^{2}>1,$ we have $%
\zeta ^{2}=0$. The numerical results of $\zeta ^{2}$ in the periodic,
quasi-periodic, and chaotic cases are illustrated in Fig.\thinspace \ref%
{SSSD-SSSB}. At first, we consider the periodic case, as shown in
Fig.\thinspace \ref{SSSD-SSSB}~(a), there is spin squeezing ($\zeta ^{2}>0$)
at the first kick, and it quickly decreases to zero. As time evolves, $\zeta
^{2}$ becomes larger than $0$. Both SSSD and SSSB appear multiple times
under this condition. In the quasi-periodic case, as shown in Fig.\thinspace %
\ref{SSSD-SSSB}(b), both SSSD and SSSB appear alternatively, but the whole
length of time intervals for zero spin squeezing parameter is longer than
that in the periodic case. At last, we consider the chaotic case, as shown
in Fig.\thinspace \ref{SSSD-SSSB}(c), the spin squeezing vanishes after a
few kicks, and then there is no spin squeezing again.

From the above discussions, we find that the quantum chaos greatly affects
the dynamics of spin squeezing and entanglement. When the initial states are
in the periodic and quasi-periodic regions, both ESD (SSSD) and ESB (SSSB)
appear alternatively, and when the initial states are in the chaotic region,
there is only ESD (SSSD).

\section{Conclusions}

In summary, we first prove that negative correlations are equivalent to spin
squeezing for systems with only symmetric Dicke states populated. Then we
study the effects of quantum chaos on spin squeezing and entanglement in the
QKT model. Using the spin squeezing parameter $\xi_T^2$, we find that, in
the periodic case the directions of the maximal spin squeezing are around
the $\vec{n_{\perp}}$-direction, while in the chaotic case, they deviate
from the $\vec{n}_{\perp}$-direction and behave irregularly. Then, we study
the dynamics of spin squeezing parameters $\xi _{KU}^{2}$, $\xi _{T}^{2}$,
and concurrence, and find that $\xi _{T}^{2}$ is a good quantity to
characterize the pairwise entanglement. At last, we study the influences of
quantum chaos on ESD (SSSD) and ESB (SSSB) in the QKT model. We find that
both ESD (SSSD) and ESB (SSSB) occur alternatively in the periodic and
quasi-periodic cases, but there is only ESD (SSSD) in the chaotic case. We
believe that the behaviors of spin squeezing in the QKT model may also be
found in other models that exhibit quantum chaos, e.g. the Dicke model.

\section*{ACKNOWLEDGMENTS}

This work is supported by NSFC with grant No.10874151, 10935010, 10947019;
and the Fundamental Research Funds for the Central Universities.

\end{document}